\documentclass{article}

\usepackage[authoryear,round]{natbib}
\usepackage[T1]{fontenc}
\usepackage[utf8]{inputenc}

\usepackage{amsmath,amsfonts,amssymb,amsthm} 

\begin{document}

\author{Tilman Sauer\thanks{tsauer@uni-mainz.de} \
and Tobias Schütz\thanks{tschuetz@uni-mainz.de}\\[0.5cm]
{\small Institute of Mathematics}\\ 
{\small Johannes Gutenberg University Mainz}\\
{\small D-55099 Mainz, Germany}}

\title{Einstein's Washington Manuscript on Unified Field Theory}

\date{Version of \today}

\maketitle

\begin{abstract}
In this note, we point attention to and briefly discuss a curious manu\-script of Einstein, composed in 1938 and entitled ``Unified Field Theory,'' the only such writing, published or unpublished, carrying this title without any further specification. Apparently never intended for publication, the manuscript sheds light both on Einstein's \emph{modus operandi} as well as on the public role of Einstein's later work on a unified field theory of gravitation and electromagnetism. 
\end{abstract}

\newpage

\section{The ``Washington manuscript''}

In July 1938, the Princeton based journal \emph{Annals of Mathematics} published a paper \emph{On a Generalization of Kaluza's Theory of Electricity} in its Vol. 39, issue No.~3 \citep{einstein_bergmann_1938}. The paper was co-authored by Albert Einstein (1879--1955) and his then assistant Peter Gabriel Bergmann (1915--2002). It presented a new discussion of an approach toward a unified theory of the gravitational and electromagnetic fields based on an extension of the number of physical dimensions characterizing space-time. Such five-dimensional theories had been discussed already a number of times, notably by Theodor Kaluza in 1921, and then again in the late twenties by Oskar Klein and others \citep{goenner_2004}. Einstein had contributed to the discussion already in 1923 and in 1927, but had given up the approach in favor of another one based on distant parallelism \citep{sauer_2014}. After giving up on distant parallelism, he had pursued another version of a five-dimensional approach, together with Walter Mayer in 1931 and 1932. The paper of 1938 indicated a return to the idea of five-dimensional theories, that he would pursue until the early forties. But he published only one other paper on the approach, this time together with Bergmann and Valentine Bargmann (1908--1989) before giving up this approach altogether for the rest of his life \citep{einstein_bargmann_bergmann_1941}.

The manuscript for the published paper was received by the \emph{Annals} on April 8, 1938,  but as correspondence between Einstein and Berg\-mann shows, corrections to the proofs were still made in late July\footnote{An English phrase of the proofs was still discussed in a letter Einstein sent to Bergmann on July 22 or 23, see Albert Einstein Archives (AEA) Call No.~6-266. In the following, our dating of archival items in the Einstein Archives is based on a reconstruction of the Einstein-Bergmann correspondence (to be published) and at times differs from the dating given in the original archival catalogue.} during a phase of intense correspondence between Einstein and Bergmann during that time.\footnote{Einstein was at Nassau Point, New York from June 15, 1938, see AEA 54-240; Bergmann was at Robinhood, Maine.} The correspondence concerned, among other things, final corrections to the proofs of this paper. In fact, in a letter to his friend Michele Besso (1873--1955) from August 8, 1938 (AEA~70-368), Einstein indicated that the paper had not been printed as late as then.

It turns out that during that same time Einstein composed another paper, which he referred to as ``my Washington manuscript'' in several letters to Berg\-mann, e.g.\ in AEA 6-271 on July 15, 1938. 
Indeed, the Library of Congress in Washington owns a 12~pp.\ holographic manuscript in Einstein's hand, signed and dated 6 July 1938.\footnote{The document carries the  Library of Congress Identification Number MSS19596.} It is written in German and entitled simply ``Unified Field Theory'' [``Einheitliche Feldtheorie'']. The Albert Einstein Archives at the Hebrew University of Jerusalem has two identical Xerox copies of a 15pp.~typescript, also entitled ``Einheitliche Feldtheorie.''\footnote{The documents carry the archival signatures AEA~2-121 and 5-008, respectively. A copy of the holograph of the Library of Congress was recently accessioned by the AEA as well and was given archival signature AEA~97-487. In addition to these complete versions of the document, AEA~2-119 is a single page with a handwritten draft version of its last paragraph, and AEA~62-789 contains a draft version of a paragraph from that document as well as pertinent calculations related to the document. AEA~62-789 is part of a batch of manuscript pages in Einstein's hand containing mainly unidentified and undated calculations, see \cite{sauer_2019}. As an example of how these unpublished working sheets can reveal Einstein's thinking and theorizing, see \cite{sauer_schuetz_2019}.}  The typescripts are typed versions of the holograph.\footnote{The transcripts contain all corrections made in the manuscript. In cases where the typed script differs from the manuscript, it was later corrected by hand to conform with the manuscript.} The handwritten equations of the typescript were added in Einstein's hand.

Our analysis of the extant documentary evidence suggests the following origin of the Washington manuscript. According to Einstein's letter of submission,\footnote{Einstein to Herbert Putnam (1861--1955), July 13, 1938 (AEA~97-494).} the manuscript was deposited with the Library of Congress at the suggestion of Einstein's friend Elias Avery Lowe (1879--1969).\footnote{Lowe's literary estate is located at 
Morgan Library in New York \citep{mayo_sharma_1990}.} Einstein and Lowe had met in 1936 when Lowe became professor for palaeography at the Institute for Advanced Study \citep{john_1994}. Lowe, however, was based in Oxford. At the same time, he held a position as reader for palaeography at Oxford where he had been teaching since 1913. When he retired as professor emeritus in Princeton in 1945, he still kept his appointment in Oxford \citep{john_1970}. Furthermore, he worked as consultant in palaeography to the Library of Congress  (AEA 97-491), \citep{john_1994}. Not only was Lowe a good friend of Einstein, but their families were friends as well, as Lowe's daughter described in her memoir of her parents \citep{lowe_2006}. In particular, Lowe's wife Helen Tracy Lowe-Porter (1877--1963),\footnote{She mainly became known as the translator of Thomas Mann \citep{romero_1980}.} whom he had married in 1911, worked as a translator for Einstein, and Einstein held her work in high esteem, as he expressed it several times in letters.\footnote{See, for instance, Einstein to Lowe, summer 1940, AEA 55-635.} In one of these letters from 1939,\footnote{(AEA 53-892), this draft of a letter was not dated by Einstein. However, he mentioned a translation regarding Gandhi's birthday, which was made in 1939, see AEA 28-459.1.} Einstein tried to advise Lowe regarding some troubles he had at the Institute.\footnote{We know from several letters (AEA 38-093, 38-094, 52-503 and 53-783) about conflicts in the Institute at the time. The Institute's faculty demanded a say regarding the succession of Flexner, who was director at the Institute from 1930 to 1939 \citep{bonner_1998}. However, it is not clear whether these are the troubles Einstein hinted at.} Einstein also spoke out for Lowe in 1944 when Lowe was about to retire, and he endorsed the latter's wish for better pension arrangements.\footnote{See Einstein's letters to Leidesdorf and to Fulton, AEA 55-632 and 55-633.} Considering the good relation to ``the Lowes'' (AEA 30-815), it seems natural that Lowe, who as a palaeograph was interested in handwritten manuscripts, suggested that Einstein donate a holograph to the Library of Congress.

In any case, Einstein complied with his wish and composed a 12pp.\ manuscript, which he signed on July 6. Before sending it off to Washington, he apparently had his secretary Helen Dukas (1896--1982) prepare a typed copy of it which he then completed by filling in the equations. The original manuscript was then sent to the Library of Congress on July 13\footnote{See his submission letter to Herbert Putnam, director of the Library of Congress (AEA 97-494).} and received there on July 19.\footnote{See the memorandum of receipt in AEA 97-493.} Einstein apparently then gave the typed version to Bergmann, in whose possession it remained.\footnote{This is indicated by Einstein in a letter to Bergmann (AEA 6-269), probably was written between July 29 and August 03, 1938.} Apparently, no copy was retained with Einstein or Dukas, as indicated by her in later years in letters to Bergmann from 1964 and 1965 (AEA 6-321 and 6-322). Bergmann, then at Syracuse University, had copies made of original Einstein documents in his possession and sent them to Helen Dukas for inclusion into the Albert Einstein Archives.\footnote{The AEA only has Xerox copies of the typescript version (AEA~2-121, 5-008), we have not been able to locate the original typescript.} After receiving the copies, Dukas first incorporated the correspondence into the archives, a process during which she learned about the existence of the so-called ``Washington manuscript.'' When she turned her attention to the unpublished scientific manuscripts, she contacted Bergmann for further information about the typescript. Upon her instigation, Bergmann, who did not remember himself any details, wrote to the Library of Congress and was informed that Einstein had indeed donated the manuscript as a gift to the Library in July 1938 (AEA 6-324 and 6-325).

\section{Why did Einstein compose the manuscript?}

We have found no evidence whatsoever that Einstein intended to publish the manuscript, even though Einstein considered it an improved presentation of the theory. The lack of any such contextual information already puzzled Dukas and Bergmann, who had no recollection of the circumstances of its composition.\footnote{In fact, Dukas even conjectured in 1964 that it might not even have been typed by herself (AEA 6-321). She changed her mind when she observed later in 1965 that the types looked like those of her old ``Remington portable'' (AEA 6-322). In her own reconstruction she also recollected that she might have been away for some time in July and August 1938. Indeed, Einstein mentioned in a letter from July 1938 that Dukas would be back on August 1 (AEA 54-583). It seems more likely to us that she did do the typing and only forgot. Since she was away in July, she could have made the transcription in a hurry which is why she may have forgotten to make a copy for her own records.} 
Given its substantial character, as we will see, the question therefore arises as to the reasons why Einstein would have written the Washington manuscript? 

We can think of three different, not mutually exclusive motivations for Einstein to compose the manuscript.

1) The submission letter  as well as the letters of acknowledgment\footnote{(AEA 97-494) and letters from Chief Assistant Librarian Martin A. Roberts (1875-1940) to Einstein (AEA 97-491) and to Lowe (AEA 97-489).} mention that Einstein wrote the manuscript and donated it to the Library of Congress at the suggestion of Lowe. Lowe was indeed acting as a consultant for paleography for the Library, and may have been involved in the creation of a collection of autographs. We know from other examples that Einstein was very willing to comply with similar requests. Already in 1924, the \textit{Lautabteilung an der Preu{\ss}ischen Staatsbibliothek}, for instance, recorded Einstein's voice for historical interest as part of a project that collected voices of famous people.\footnote{Einstein's lost record carries archival number Aut 56 and identification number 16189, see https://www.lautarchiv.hu-berlin.de/objekte/lautarchiv/16189/ (retrieved on February 26, 2020). See \citep[Doc.~208]{cpae14}, for the text and further context. For an even earlier voice recording, see \citep[Vol.7, Doc.50a]{cpae13}.} Furthermore, Einstein was interested in the study of handwriting, as a visit to a graphologist in 1930 shows \citep{modern_review_1930}.

Although we have found no evidence for it in the case of this particular donation, such autograph donations were also serving, both before and after 1938, specific purposes of fund raising. For instance, Einstein donated, after some dispute with the astronomer Erwin Freundlich (1885--1964), a manuscript of his publication ``The Foundation of the General Theory of Relativity'' (``Die Grundlage der allgemeinen Relativitätstheorie'') from 1916 \citep{einstein_1916} to the \textit{Jewish National and University Library} in 1925 in order to support the university library as well as other charities \citep{gutfreund_renn_2015}. A similar donation by Einstein happened in 1943. In order to promote the sale of war bonds, Einstein prepared a handwritten copy of his famous special relativity paper ``On the electrodynamics of moving bodies'' (``Zur Elektrodynamik bewegter Körper'') \citep{einstein_1905}.\footnote{See AEA 5-025 for a copy of that manuscript.} This manuscript was then auctioned and presented by the buyer to the Library of Congress. Again, a later manuscript on unified field theory entitled ``the bi-vector field'' (``Das Bi-Vektor Feld'') was auctioned for the same purpose and given to the Library in late 1944 or early 1945 \citep{loc_aep,brasch_1945}.

2) Abraham J.Karp (1921--2003) suggested that the donation was given by Einstein as a sign of gratitude to the United States \citep{karp_1991}. Given Einstein's situation this appears quite possible. He was himself in the process of naturalization, having applied in May 1935 for US citizenship, which was then obtained in 1940 \citep{calaprize_kennefick_schulmann_2015}.  He was also very active in helping friends and colleagues who were suffering persecution in Nazi Germany to emigrate to the United States. Karp does not provide any further evidence to support his conjecture. On receiving the manuscript, the Library of Congress issued a press release which, however, mentioned or even only suggested no such connection.\footnote{See AEA 97-492, for a draft of the press release. We have not found evidence for reception of the press announcement.} 

3) Einstein may also have welcomed the writing of a holograph for the Library as occasion for a moment of reflection. This interpretation is suggested by Einstein's comments about the manuscript to Bergmann. There he emphasized that he had ``reworked the whole theory anew'' and ``presented the new theory systematically'' and found that ``the whole thing now takes on a really beautiful form, and I really have joy with it.''\footnote{Einstein to Bergmann in AEA~6-256 on July 12, 1938. See also AEA 6-271.} This interpretation is also suggested by a closer look at the manuscript itself.

\section{Characteristics of the manuscript}

The substance of the Washington manuscript repeats what is contained already in the published Einstein-Bergmann paper. But there are significant differences.

The essence of the Einstein-Bergmann paper is to extend Kaluza's original five-dimensional theory \citep{Kaluza_1921} by giving the fifth dimension reality, and at the same time replacing its so-called cylinder condition by a periodicity condition. On the basis of this ``generalization,'' Einstein and Bergmann then derived the following field equations
\begin{align}
& \underset{\mathrm{1}}{\alpha} \left( \frac{1}{2} R_\mathrm{kl} + \frac{1}{2} R_\mathrm{lk} - \frac{1}{2} g_\mathrm{kl} R \right) + \underset{\mathrm{2}}{\alpha} \left( 2 A_\mathrm{km} A_\mathrm{l}^\mathrm {\hphantom{l}m} - \frac{1}{2} g_\mathrm{kl} A_\mathrm{mn} A^\mathrm {mn} \right) \notag \\
&+ \underset{\mathrm{3}}{\alpha} \left(- 2 \partial_\mathrm{0} \partial_\mathrm{0} g_\mathrm{kl} + 2 g^\mathrm {rs} \partial_\mathrm{0} g_\mathrm{kr} \partial_\mathrm{0} g_\mathrm{ls} - g^\mathrm {rs} \partial_\mathrm{0} g_\mathrm{rs} \partial_\mathrm{0} g_\mathrm{kl} - \frac{1}{2} \partial_\mathrm{0} g^\mathrm {rs} \partial_\mathrm{0} g_\mathrm{rs} g_\mathrm{kl}  \right) \notag \\
&+ \underset{\mathrm{4}}{\alpha} g_\mathrm{kl} \left( \frac{1}{2} \left( g^\mathrm {mn} \partial_\mathrm{0} g_\mathrm{mn} \right)^2 + 2 g^\mathrm {mn} \partial_\mathrm{0} \partial_\mathrm{0} g_\mathrm{mn} + 2 \partial_\mathrm{0} g^\mathrm {mn} \partial_\mathrm{0} g_\mathrm{mn} \right) = 0,
\end{align}
and
\begin{align}
\displaystyle\int \left( \underset{\mathrm{1}}{\alpha} \left( g^\mathrm {mn} \partial_\mathrm{0} \Gamma^{\mathrm{s}}_{\mathrm{mn}} - g^\mathrm {ms} \partial_\mathrm{0} \Gamma^{\mathrm{n}}_{\mathrm{mn}} \right) - 4 \underset{\mathrm{2}}{\alpha} \nabla_\mathrm{t} A^\mathrm {st} \right) \sqrt{-g} \,d x^\mathrm 0= 0
\end{align}
for a five-dimensional theory, derived from a variational principle \citep{einstein_bergmann_1938}. In these equations, the $g_{\mathrm{ab}}$ are components of a five-dimensional metric tensor, which effectively reduces to a four dimensional one since its components with index zero vanish. The difference between Kaluza's theory and that of Einstein and Bergmann lies in the fact that the components of this metric are now periodic functions of $x^{\mathrm{0}}$, the fifth coordinate. This also gives rise to the remaining integral in the second equation.  The quantities $A_{\mathrm{mn}}$ are the antisymmetrized derivatives of the vector field $A_{\mathrm{m}}$, which was interpreted as representing the electric potential. $R_{\mathrm{kl}}$ and $R$ are the (5-d) Ricci tensor and Ricci scalar, while $\Gamma^{\mathrm{s}}_{\mathrm{mn}}$ (appearing only in the second equation) denotes the (5-d) Christoffel symbol. The quantities $\underset{\mathrm{i}}{\alpha}$ are constants. A discussion of these constants was given by Einstein in a later unpublished manuscript.\footnote{See ``Ein Gesichtspunkt für eine spezielle Wahl der in der verallgemeinerten Kaluza-Theorie auftretenden Konstanten'' (AEA 1-136), dated to 1941 by the Albert Einstein Archives catalogue. See also \citet{dongen_2010}.}

As regards a unified field theory based on these field equations, the Washington manuscript contains nothing new, but the differences between the manuscript and the published paper are nevertheless interesting and revealing.

A most obvious difference between the published paper and the Washington manuscript concerns the title and authorship. Given that the manuscript did not contain anything new of substance and Einstein's and Bergmann's cooperation was ongoing, one might have expected that Bergmann would have been co-author here as well. Instead, we learn from the correspondence that Einstein did not even bother to consult with Bergmann about this manuscript. Equally telling perhaps is the change in title. While the title of the published paper is a fairly technical one, alluding to Kaluza's earlier theory, the manuscript is simply entitled ``Unified Field Theory.'' To the best of our knowledge, this is the only document, published or unpublished, by Einstein that is entitled ``unified field theory'' without any further qualification.\footnote{Similar titles containing the phrase ``Unified Field Theory'' include a 1925 paper on the metric-affine approach, entitled ``Einheitliche Feldtheorie von Gravitation und Elektrizit\"at'' \citep{einstein_1925a}, or a paper with his assistant Mayer on another variant of the Kaluza-Klein approach, entitled ``Einheitliche Theorie von Gravitation und Elektrizit\"at" \citep{einstein_mayer_1931,einstein_mayer_1932}. Perhaps closest to the manuscript is a paper entitled ``Zur einheitlichen Feldtheorie'' \citep{einstein_1929a} or a French paper ``Théorie unitaire du champ physique'' \citep{einstein_1930}. The latter two items are from his teleparallel approach. Interestingly, the 1929 paper was the one that created out-of-proportion public attention when it came out in 1929, as a result of international media coverage, see \citep[p. 414]{sauer_2006} and \citep{sauer_2014}.} Both the change in authorship and the change of title are compatible with the donation request as the primary motivation for its composition. If Einstein never intended to publish this manuscript, he may instead have wanted to comply with Lowe's wish of having an autograph manuscript and the Library's interest in obtaining something that could be put on display to a larger public.

Notwithstanding the fact that Bergmann no longer figured as a co-author, Einstein made the latter's  co-responsibility clear in the introductory paragraph of the manuscript which also makes explicit what he intended to be the new feature of the manuscript. The introductory paragraph reads:
\begin{quote}
In the last months, I have developed, together with my assistant P.~Bergmann, a unified field theory, which emerged as a generalization of Kaluza's theory of the electric field. In the following this theory shall be presented independently from its historical roots, in order that its logical structure may come to the fore as clearly as possible.\footnote{``In den letzten Monaten habe ich zusammen mit meinem Assistenten P. Bergmann eine einheitliche Feldtheorie entwickelt, welche durch Verallgemeinerung von Kaluza's Theorie des elektrischen Feldes entstanden ist. Im Folgenden soll diese Theorie unabh\"angig von ihren historischen Wurzeln dargestellt werden, damit ihre logische Struktur m\"oglichst deutlich hervortrete.'' (AEA 2-121, 97-487).}
\end{quote}

That Einstein presented their theory independently from Kaluza's theory is one of the main differences to the publication. There, Einstein and Bergmann first gave a recapitulation of Kaluza's theory. They then illustrated how to alter Kaluza's theory in order to ascribe a ``physical reality to the fifth dimension'' \citep[p.~683]{einstein_bergmann_1938} and, therefore, they extended Kaluza's theory. In Einstein's Washington manuscript, on the other hand, the theory was developed from scratch on the basis of independent axioms.

The theory was now exclusively based on three axioms. The first one postulated a five-dimensional space, equipped with a regular Riemannian metric.
\begin{quote}
A five-dimensional space with a regular Riemann-metric
$$
d\sigma^2=\gamma_{\mu\nu}dx_{\mu}dx_{\nu}
$$
is assumed to be the basis for the theory.\footnote{``Es wird ein fünf-dimensionaler Raum mit einer regulären Riemann-Metrik $d\sigma^2=\gamma_{\mu\nu}dx_{\mu}dx_{\nu}$ zugrunde gelegt.''}
\end{quote}
The axiom further stipulated that the metric could locally be transformed to a diagonal metric of the form $d\sigma^2=dx_1^2+dx_2^2+dx_3^2-dx_4^2+dx_0^2.$

The second axiom introduced spatial compactification of the additional dimension by requiring periodicity along the fifth coordinate.
\begin{quote}
With respect to the dimension characterized by the coordinate $x_0$, the space be closed.\footnote{``Bezüglich der durch die Koordinate $x_0
$ charakterisierten Dimension sei der Raum in sich geschlossen.''}
\end{quote}
This requirement was then further concretized by stipulating the possibility of finding coordinates in which the $\gamma_{\mu\nu}$ are periodic functions of $x_0$.

The third axiom introduced a congruence of geodesics by requiring the existence of a unique, singularity-free, closed space-like geodesic through each point.
\begin{quote}
Through each point of our space there shall exist one and only one geodesic line that is closed without singularities and ``space-like''.\footnote{``Durch jeden Punkt unseres Raumes soll es eine und nur eine in sich singularitätsfrei geschlossene ``raumartige'' geodätische Linie geben.''}
\end{quote}
Again this point was concretized by stipulating that in the periodic representation the geodesic should be unique and pass through all periodically repeated points.

By these three axioms, Einstein wrote, ``the space structure, which underlies the theory, is characterized completely'' (``Damit ist die Raumstruktur, welche der Theorie zugrundeliegt, vollständig charakterisiert.'')

Einstein here emphasizes the fact that the substance of the theory is captured in just three axioms. This explicitness is remarkable in view of his earlier methodological reflections on the axiomatic method. The axiomatic formulation of the Washington manuscript is clearly an example of his understanding that
\begin{quote}
[t]he goal of theoretical physics is to create a logical system of concepts based on the fewest possible mutually independent hypotheses, allowing a causal understanding of the entire complex of physical processes, 
\end{quote} 
as he expressed it in 1922 in reflections ``on the present crisis of theoretical physics'' \cite[p.~1]{einstein_1922}.
Nevertheless, Einstein's attitude towards the role of axioms has been somewhat ambivalent ever since his experience of competitive efforts in completing the general theory of relativity with David Hilbert (1862--1943) in 1915. While Hilbert made the use of the axiomatic method the hallmark of his heuristics, Einstein acknowledged its use only somewhat hesitantly. In his own exposition of the new theory of general relativity in 1916, he emphasized that it was \emph{not} his purpose 
\begin{quote}
to represent the general theory of relativity as a system that is as simple and logical as possible, and with a minimum number of axioms;
\end{quote}
and that instead his aim was to 
\begin{quote}
develop this theory in such a way
that the reader will feel that the path we have entered upon is psychologically the natural one, and that the underlying assumptions will seem to have the highest possible degree of security. \citep[p.~777]{einstein_1916}
\end{quote}
And in contemporary correspondence, Einstein expressed himself rather critical about the axiomatic \emph{method}, which he felt could not help anything in finding the suitable hypotheses.\footnote{See, e.g., his letter to Hermann Weyl, 23 November 1916 \cite[Doc.~278]{cpae08}.} Nevertheless, an axiomatic formulation of a physical theory seems to have been his ultimate goal, and, as an example, in his discussion of Eddington's unified theory based on a general affine connection he was ready to refer to the variational principle underlying the theory as an axiom, just as Hilbert had done \citep[p.~367]{einstein_1925b}. The transition from a genetic exposition of explaining the difficulties of the earlier Kaluza-Klein theory as a justification for the new ansatz toward a purely axiomatic presentation again reflects this tension of the role of axiomatics in Einstein's thinking.

The second main difference from the publication was pointed out by Einstein in a letter to Bergmann (AEA 6-256), when he noted that he incorporated the mathematical appendix of their publication into the main text. Instead of treating tensor densities and the mathematical part of the derivation of the field equations separately, Einstein now incorporated this part into the main text.\footnote{In fact, Einstein's manuscript does not have an appendix.} In another letter to Bergmann (AEA 6-271), Einstein noted that, regarding the structure, he preferred his manuscript over their publication.

When Einstein and Bergmann first considered Kaluza's theory and then generalized it, they also considered first the classical theory of general relativity in their appendix by introducing tensor densities and by deriving the field equations for the four-dimensional case. They then transferred the procedure to the five-dimensional generalized Kaluza theory. But just as Einstein developed the theory independently from Kaluza's theory in his manuscript, he also did not consider the four-dimensional theory first. Instead he  started axiomatically with the new five-dimensional theory as an independent theory. 

In addition to the above mentioned differences, the two works also differ with respect to minor aspects, as regards e.g.\ the notation. For instance, instead of denoting the equivalent to the electric potential by $A_{\mathrm{m}}$, he used the notation $\varphi_{\mathrm{m}}$. This change in notation apparently created some confusion for Einstein himself, when he miswrote $A_{\mathrm{mn}}$ instead of $\varphi_{\mathrm{mn}}$ in his manuscript.\footnote{See page 9 of his holographic version AEA 97-487. The confusion about the notations also appears in his correspondence with Bergmann, see his letter AEA 6-266.}

In summary, the Washington manuscript, as we have seen, essentially gave another derivation of the field equations of the Einstein-Bergmann paper but based the discussion on a more logical foundation and proceeded in a direct mathematical way which does not relegate mathematical details into an appendix. It also deleted the historical origin of the Einstein-Bergmann theory as an extension of Kaluza's theory in favor of a purely axiomatic foundation. In doing so, Einstein decided to figure as the sole author of this new presentation of the theory, and he gave the manuscript the unusually generic title of ``unified field theory'' without any further qualification.

\section{Concluding remarks}

In contrast to the published Einstein-Bergmann paper which does not indicate in which direction Einstein and Bergmann wanted to proceed further, the Washington manuscript ends with the following paragraph:
\begin{quote}
The theory developed here provides a unified conception of the structure of physical space which is completely satisfactory from a formal point of view. Further investigations will have to show whether it contains a theory (free of statistical elements) of elementary particles and of the quantum phenomena.\footnote{``Die im Vorstehenden entwickelte Theorie gibt eine formal v\"ollig befriedigende einheitliche Auffassung von der Struktur des physikalischen Raumes. Weitere Untersuchungen m\"ussen zeigen, ob sie eine (von statistischen Elementen freie) Theorie der Elementar-Teilchen sowie der Quanten-Ph\"anomene enth\"alt.'' (AEA 2-121, 97-487).}
\end{quote}
Three years later, in 1941, Einstein and Bergmann, now together with Valentine Bargmann, published a follow-up  to the first Einstein-Bergmann paper \citep{einstein_bargmann_bergmann_1941}. In it, they began with a brief description of the earlier theory, going back in spirit, if not in exact phrasing, to Einstein's Washington manuscript, characterizing the core of the theory in terms of three ``axioms.'' But this paper by Einstein, Bargmann, and Bergmann was Einstein's last published attempt along higher-dimensional approaches to unified field theory. Their paper ends, much more sceptically, by listing their reasons for the non-viability of the five-dimensional approach. Two years later, \citet{einstein_pauli_1943} closed the lid on this approach by publishing a proof of ``the non-existence of regular stationary solutions of relativistic field equations'' in both four and five dimensions. As \citet{dongen_2002,dongen_2010} has already emphasized, foremost among the reasons for Einstein's giving up on the five-dimensional approach figured their failure to find particle-like solutions, a task which had been formulated at the end of the Washington manuscript, but not in the published paper. The failed attempt to find particle solutions are documented only in working sheets and unpublished calculational notes \citep{sauer_2019}. This fact and the characteristics of the Washington manuscript show the necessity of including unpublished notes and correspondence for a proper historical understanding of the unified field theory program.

Despite the highly technical character of Einstein's manuscript which would have made it understandable only for a handful of contemporaries, the holograph donation to the Library of Congress also documents the enduring fascination of the public with Einstein and his unified field theory program at the eve of the Second World War, a fascination that Einstein apparently was willing to play along with to some extent, despite his primary interest in the technical details of the program itself.

\end{document}